\input{aipcheck}

\documentclass[
    ,final            
  ]
  {aipproc}

\layoutstyle{6x9}

\newcommand{\1}{1\!\!\!\bot}

\newcommand*{\bea}{\begin{eqnarray}}
\newcommand*{\eea}{\end{eqnarray}}

\begin{document}

\title{Infrared Maximally Abelian Gauge}


\author{Tereza Mendes}{
}

\author{Attilio Cucchieri}{
}

\author{Antonio Mihara}{
  address={Instituto de F\'\i sica de S\~ao Carlos, Universidade de S\~ao Paulo, 
Caixa Postal 369, \\ 13560-970 S\~ao Carlos, SP, Brazil}
}

\begin{abstract}
The confinement scenario in
Maximally Abelian gauge (MAG)
is based on the concepts of Abelian dominance and of dual superconductivity.
Recently, several groups pointed out the possible existence in MAG
of ghost and gluon condensates with mass dimension 2,
which in turn should influence the infrared behavior of ghost and gluon
propagators.
We present preliminary results for the first lattice numerical study
of the ghost propagator and of ghost condensation
for pure $SU(2)$ theory in the MAG.
\end{abstract}

\maketitle

The study of the infra-red (IR) limit of QCD is of central
      importance for understanding the mechanism of
      confinement.
Despite being non-gauge-invariant,
      gluon and ghost propagators are powerful
      tools in the (non-perturbative) investigation of this limit.
In recent years, (gauge-dependent) condensates of mass
      dimension two have received considerable attention.
An example of such objects is the ghost condensate \cite{Schaden:1999ew,Capri:2005tj},
      related to the breakdown of a global $SL(2,R)$ symmetry. 
In particular, in MAG the diagonal and
      off-diagonal components of the ghost propagators
      are expected to be modified by ghost condensation.
In this paper we present preliminary results of lattice
studies of the Faddeev--Popov (FP) matrix for pure $SU(2)$ theory 
in the MAG. We consider the ghost propagator, the ghost condensate
and the smallest eigenvalue of the FP matrix.

On the lattice, for the $SU(2)$ case, the MAG is obtained (see e.g.\ 
\cite{Bornyakov:2003ee}) by minimizing the functional
\begin{equation}
S \;=\; - \frac{1}{2 d V} \sum_{x,\mu} Tr
\left[ \sigma_3 U_{\mu}(x) \sigma_3 U^{\dagger}_{\mu}(x) \right]\,.
\end{equation}
At any local minimum one has that the Faddeev-Popov matrix, defined as
\begin{eqnarray}
\!\!\!\!\!\!\!\! \sum_{b y} M^{ab}(x,y) \gamma^b(y) \!\!\!\! &=& \!\!\!\! \sum_{\mu}
\gamma^a(x) [ V_{\mu}(x) + V_{\mu}(x-e_{\mu}) ]
          \, + \,2 \, \{ \gamma^a(x-e_{\mu}) [ 1 - 2 (U_{\mu}^0(x))^2] 
               \nonumber \\[3mm]
\!\!\! & & \!\!\! - \, 2\, \sum_{b} \gamma^b(x-e_{\mu}) [ \epsilon_{ab} U_{\mu}^0(x) U_{\mu}^3(x) +
           \sum_{cd} \epsilon_{ad} \epsilon_{bc} U_{\mu}^d(x) U_{\mu}^c(x) ] \}
\, ,
\end{eqnarray}
is positive-definite. Here the color indices take values
$1, 2$ and we follow the notation
$U_{\mu}(x) = U_{\mu}^0(x) \1 + i\, {\sigma}^a {U}_{\mu}^a(x)$ and
$V_{\mu}(x) = (U_{\mu}^0(x))^2+(U_{\mu}^3(x))^2-(U_{\mu}^1(x))^2-(U_{\mu}^2(x))^2$,
where $\sigma^a$ are the 3 Pauli matrices.
Notice that (as in Landau gauge \cite{Cucchieri:2005yr}) this matrix
is symmetric under the simultaneous exchange of color and space-time
indices.
Using the relation $U_{\mu}(x) = \exp{[- i a g_0 A_{\mu}(x)]}$
one finds (in the formal continuum limit $a \to 0$) the standard
continuum results \cite{Bruckmann:2000xd} 
for the stationary conditions above and for $M^{ab}(x,y)$.

We have considered four values of $\beta$ (2.2, 2.3, 2.4, 2.512)
and lattice volumes up to $40^4$.
Our results for the gluon propagators are in agreement with the 
study by Bornyakov et al.\ \cite{Bornyakov:2003ee}:
we see a clear suppression of the off-diagonal propagators compared to
the diagonal (transverse) one, supporting Abelian dominance.
We have fitted our data for the various gluon
propagators (at all values of $V$ and $\beta = 2.2$),
obtaining the following behaviors. For $D(p^2)$ (transverse) diagonal,
our data favor a Stingl-Gribov form 
\begin{equation}
D(p^2) \;=\; \frac{1+d\,p^2}{a+b\,p^2+c\,p^4}\,,
\label{SG}
\end{equation}
with a mass $\,m = \sqrt{a/b} \,\approx \, 0.72\,GeV\,$.
Note that the above equation corresponds to a pair
of complex conjugate poles $z$ and $z^*$. We can thus write
$z = x + i y$ with $x = b/(2 c) \approx 0.32 \, GeV^2$ and
$y = \sqrt{a/c - x^2} \approx 0.47 \, GeV^2$. Let us recall that
in the case of a Gribov-like propagator these two poles are
purely imaginary.
%
For $D(p^2)$ transverse off-diagonal our best fit is
of Yukawa type, i.e.\ $D(p^2) = 1/(a+b\,p^2)\,$, with
a mass $\,m = \sqrt{a/b} \,\approx \, 0.97\,GeV\,$.
Finally, the longitudinal off-diagonal gluon propagator
is best fitted by
$D(p^2) = 1/(a+b\,p^2+c\,p^4)$ (i.e.\ also of Yukawa type) 
with a mass $\,m = \sqrt{a/b} \,\approx \, 1.25\,GeV\,$.
As expected from Abelian dominance, the mass is larger
in the off-diagonal case.

\begin{figure}
\includegraphics[width=0.45\textwidth]{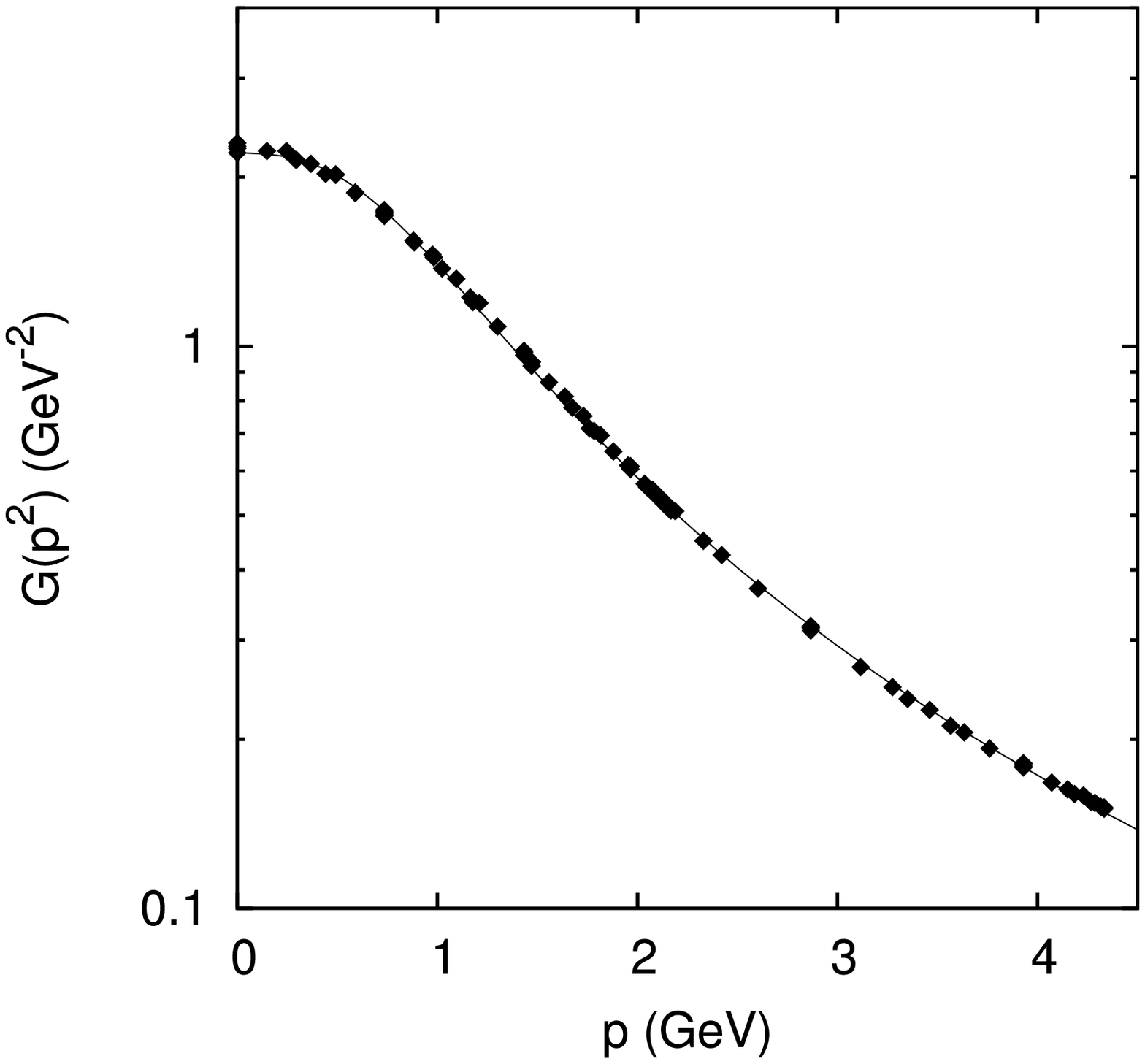} \hspace{5mm}
\includegraphics[width=0.48\textwidth]{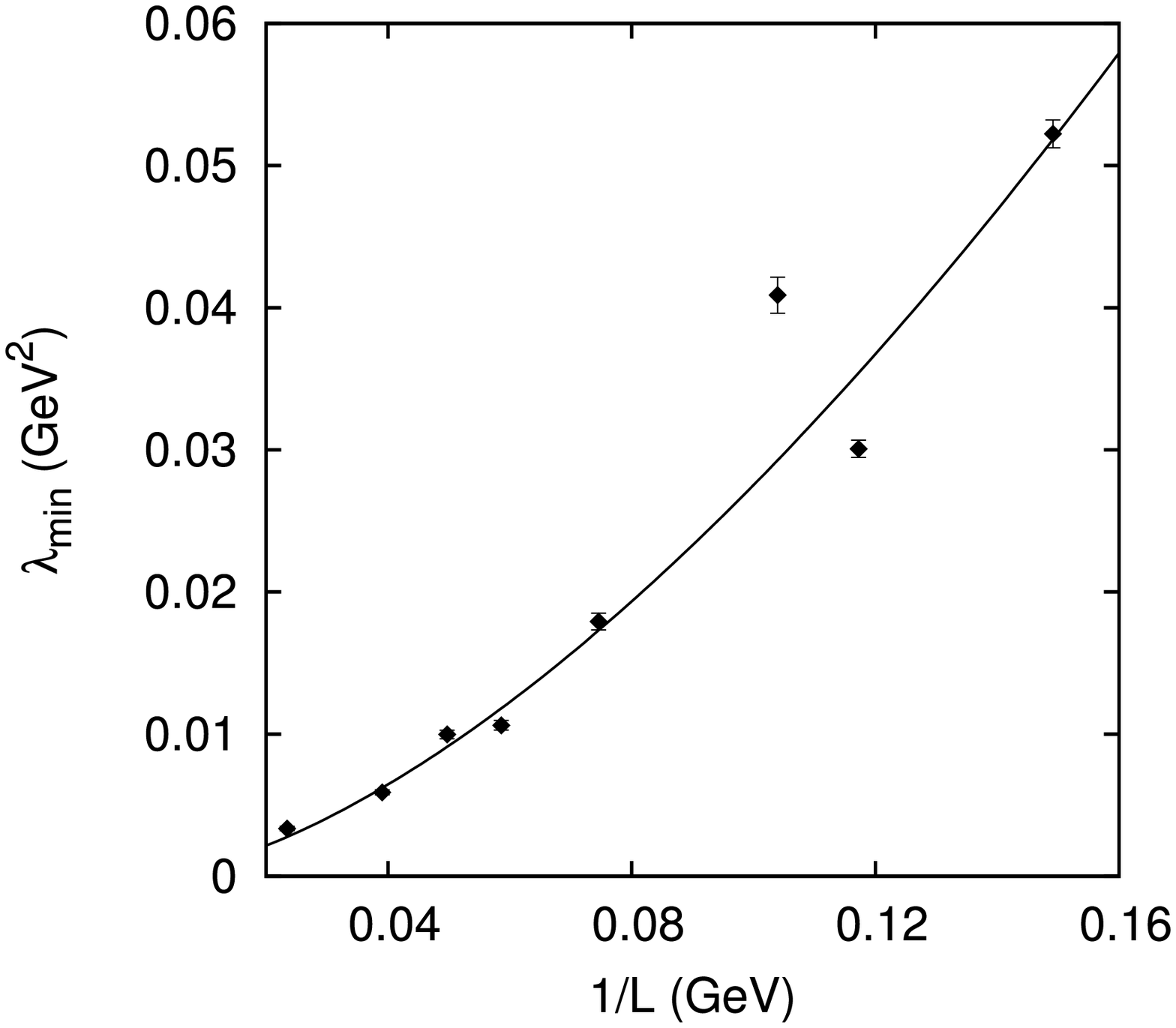}
\caption{Left: plot of $G(p^2)$ as a function of improved $p$
for lattice volumes $V = 16^4$, $24^4$, $40^4$ and $\beta = 2.2$.
Right: plot of the smallest eigenvalue of the FP operator, as a
function of the inverse linear size of the system.
}
\label{ghost}
\end{figure}

In Fig.\ \ref{ghost} (left) we show our data for the 
ghost propagator $G(p^2)$, as a function of an improved momentum $p$
(see Ref.\ \cite{Ma:1999kn}).
The data show little volume dependence at small $p$.
(Note that, contrary to Landau gauge, here we can evaluate the ghost 
propagator at zero momentum.)
We see no sign of an enhanced IR propagator.
We have fitted our data (at $\beta = 2.2$),
obtaining a behavior of the type (\ref{SG}) above
with
$a = 0.45(1)\,GeV^2\,,$
$b = 1.1(3)\,,$
$c = 0.73(30)\,GeV^{-2}\,,$
$d = 2.1(9)\,GeV^{-2}\,.$
Thus, we see a Stingl-Gribov fit with mass
$\,m \,\approx \, 0.6\,GeV\,$ and complex poles given by
$x\approx 0.75$, $y\approx 0.22$.

We next consider (see Fig.\ \ref{ghost}, right)
the smallest eigenvalue of the FP matrix.
We have looked at $\lambda_{min}$ for several lattice volumes and values 
of $\beta$ as a function of $1/L$. The data are fitted
to $\,a\,(1/L)^b\,$ with $\,b = 1.6(1)$, showing that $\lambda_{min}$
vanishes more slowly than $(1/L)^2$ (Laplacian).
This may explain why we do not see a diverging ghost propagator at zero
momentum even at rather large lattice volumes \cite{Cucchieri}.

\begin{figure}
\includegraphics[width=0.45\textwidth]{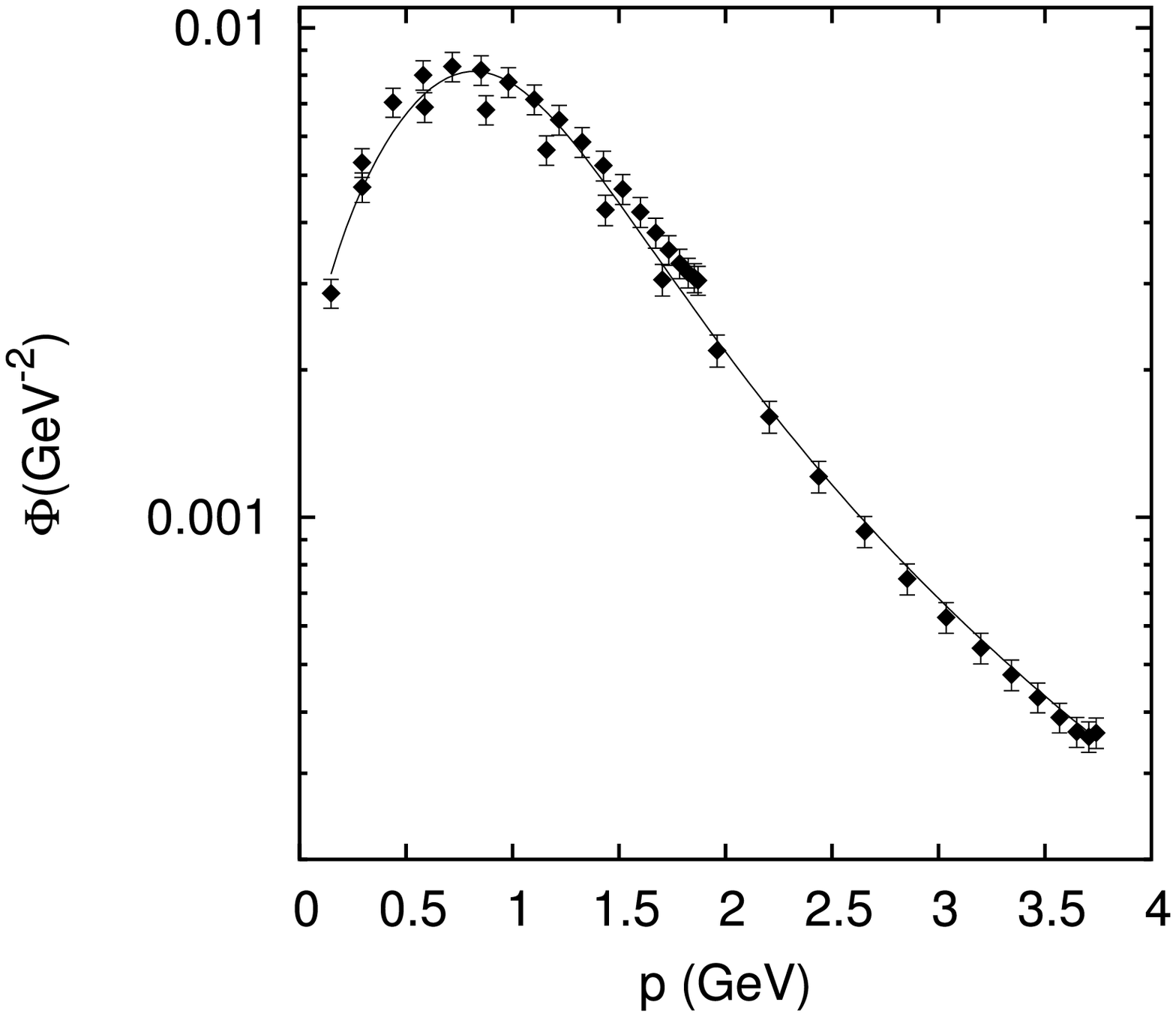} \hspace{5mm}
\includegraphics[width=0.45\textwidth]{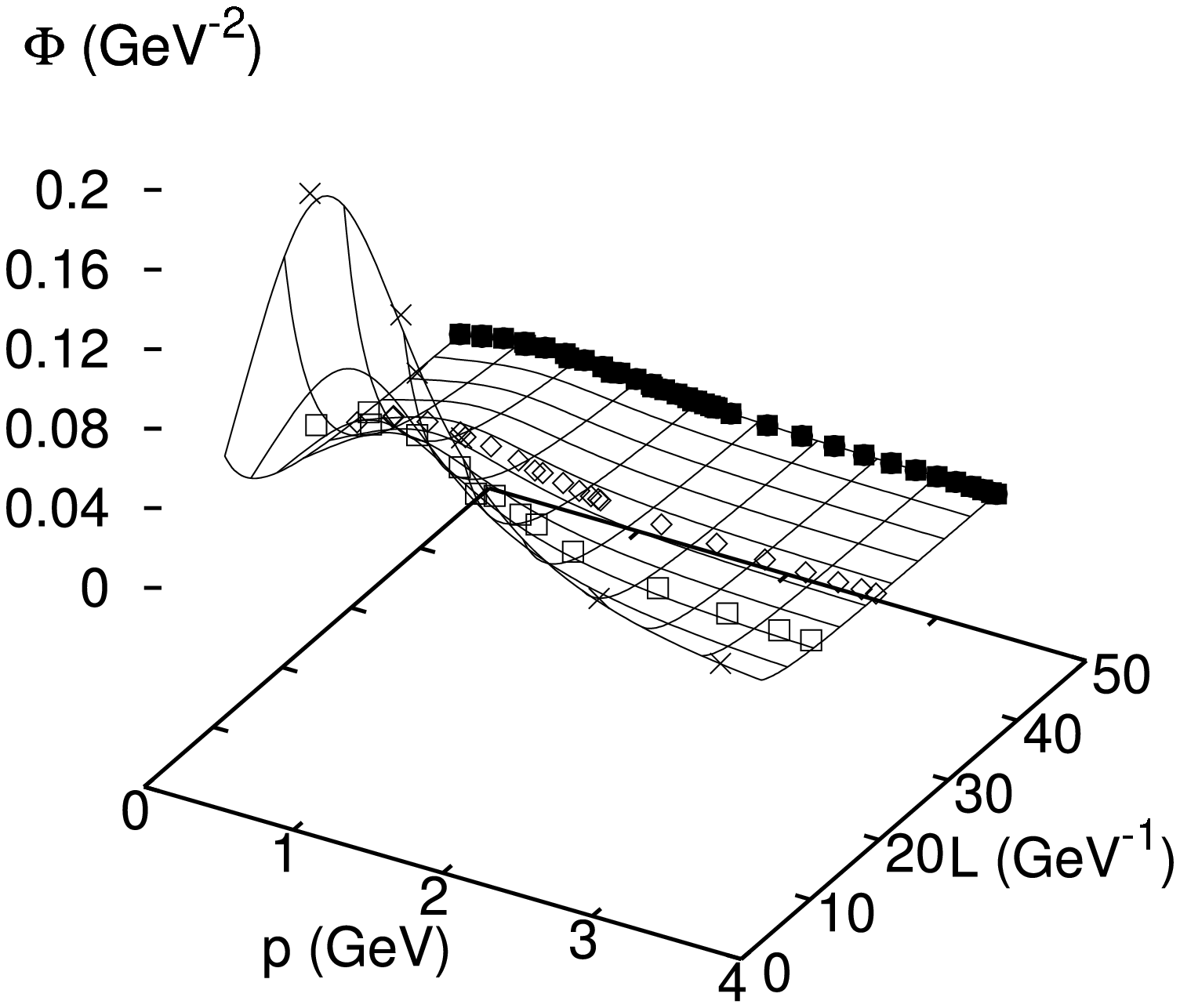}
\caption{Left: plot of the quantity $\Phi(p^2) =
L^2 / \cos\left(\pi \,\widetilde{p}_{\mu}\, a/L\right)
\langle \, | \, \epsilon_{ab} G^{ab}(p^2) /2 \, | \,\rangle $
as a function of $p$ for lattice
volumes $V = 8^4$, $16^4$, $24^4$, $40^4$ and $\beta = 2.2$.
Right: plot of $\Phi(p^2)$ as a function of $p$ and $L$.}
\label{cond}
\end{figure}

Following the analysis done in Landau gauge \cite{Cucchieri:2005yr}, 
we consider the anti-symmetric off-diagonal ghost propagator
$ \langle \, | \, \epsilon_{ab} G^{ab}(p^2) /2 \, | \,\rangle $
rescaled by $L^2 / \cos\left(\pi \,\widetilde{p}_{\mu}\, a/L\right)$,
as a function of the (unimproved) momentum $p$ for all lattice volumes 
and $\beta$ values considered.
The data show nice scaling for all cases considered.
The data at $\,V = 40^4$ and $\,\beta = 2.2\,$ can be fitted by
$\,\Phi(p) = (a+b\,p/L^2)(p^4+v^2)\,$ with
$a = 0.0026(7)\,GeV^2\,,$
$b = 32.6(7)\,GeV^{-1}\,$
and
$v^2 = 1.7(1)\,GeV^{4}\,.$
We thus have a rather large ghost condensate
$\,v \approx 1.3 \,GeV^2\,$, but we cannot be sure that it survives
in the infinite-volume limit, since the overall constant $a$ might be null.
We can also fit data at several $V$'s and $\,\beta$'s for
$\,\Phi(p^2)\,$ as a function of $p$ and $L$ (see Fig.\ \ref{cond}, right).
We obtain $\Phi(p) = (a+b\,p/L^2)(p^4+v^2)\,$ with
$a = 0.0033(6)\,GeV^2\,,$
$b = 35.8(5)\,GeV^{-1}\,$
and
$v^2 = 1.87(8)\,GeV^{4}\,.$
We note that the fit parameters change little with the (physical)
lattice volume.

We are currently investigating the effects of Gribov copies on our results.

\vskip 3mm
We thank M.I. Polikarpov and M. Schaden for helpful 
discussions. A.\ C.\ and T.\ M.\ were supported by FAPESP 
and CNPq. A.\ M.\ was supported by FAPESP.

\end{document}